# Coupled-mode theory for stimulated Raman scattering in high-$Q/V_m$ silicon photonic band gap defect cavity lasers

Xiaodong Yang and Chee Wei Wong

Optical Nanostructures Laboratory, Columbia University, New York, NY 10027

We demonstrate the dynamics of stimulated Raman scattering in designed high-$Q/V_m$ silicon photonic band gap nanocavities through the coupled-mode theory framework towards optically-pumped silicon lasing. The interplay of other $\chi^{(3)}$ effects such as two-photon absorption and optical Kerr, related free-carrier dynamics, thermal effects, as well as linear losses such as cavity radiation and linear material absorption are included and investigated numerically. Our results clarify the relative contributions and evolution of the mechanisms, and demonstrate the lasing and shutdown thresholds. Our studies illustrate the conditions for continuous-wave and pulsed highly-efficient Raman frequency conversion to be practically realized in monolithic silicon high-$Q/V_m$ photonic band gap defect cavities.

## 1. Introduction

Silicon is increasingly being considered as the dominant platform for photonic integrated circuits, providing the advantages of dense CMOS electronics integration, transparency in the telecommunication C-band, as well as high-index contrast for compact integrated optical functionalities. Passive silicon photonic devices have already recently reported remarkable progress [1-4]. With strong modal confinement, nonlinear optical properties in silicon are enhanced and active functionalities in highly integrated silicon devices have been realized, such as all-optical switches [5], and all-silicon Raman amplification and lasing [6-14].

Photonic crystals (PhC), with periodic modulation in the index of refraction, permit unique opportunities in specific studies and applications [15, 16]. Two-dimensional (2D) PhC slabs confine light by Bragg reflection in-plane and total internal reflection in the third dimension. Introduction of point and line defects into 2D PhC slabs create nanocavities and PhC waveguides with arbitrary dispersion control that can be designed from first principles. Such defect cavities have subwavelength modal volumes, on order $0.02(\lambda/n)^3$, corresponding to high field intensities per photon for increased nonlinear interaction. Through $k$-space design of cavity modes [17, 18], recently cavities with remarkable high quality factors ($Q$) [1, 2, 19-22] have been achieved, permitting for long photon lifetimes for light-matter interactions. The strong optical confinement and photon lifetimes in these cavities permit fundamental studies and integrated nanophotonics applications [23], such as channel add/drop filters [24], low-threshold quantum well lasers [25], cavity quantum electrodynamics [26], enhancement of optical nonlinearities [27], and ultrasmall nonlinear bistable devices [28, 29].

Raman scattering in silica-based high-$Q$ microcavities, such as microspheres [30], microdisks and microtoroids [31], have shown remarkable ultra-low lasing thresholds. In addition, Raman lasing in silicon waveguides has also been observed [6-14] where the bulk Raman gain coefficient $g_R$ is $10^3$ to $10^4$ times larger in silicon than in silica and two-photon absorption induced free-carrier absorption [32] addressed by pulsed or reversed biased p-i-n diode operation. To achieve significant amplification and ultimately lasing, the gain medium should be placed in a cavity with sufficiently high-$Q$, and ultrasmall modal volumes $V_m$. The enhanced stimulated Raman amplification and ultralow threshold Raman lasing in high-$Q/V_m$ photonic band gap nanocavities was suggested [33]. Stimulated Raman scattering (SRS) in periodic crystals with slow group velocity was theoretically studied with a semiclassical model [34], and enhancement in line-defect photonic crystal slow-light waveguides was also proposed [35].

Here we employ a coupled-mode theory framework to study the various contributions on Raman scattering and lasing [36, 37]. Coupled-mode equations are widely used in passive photonic devices such as optical waveguide direction couplers, channel add-drop filters [38, 39] and also in the analysis of optical nonlinearities [28, 40-42]. The coupled-mode equations for stimulated Raman scattering in silicon-on-insulator waveguides [9, 43-45], fiber Bragg grating [46] and silica microsphere [47, 48] have been studied. Following from the work of Ref [28, 40, 42], we derive, in this present paper, the coupled-mode equations for stimulated Raman scattering in high-$Q/V_m$ silicon photonic band gap nanocavities. The dynamics of coupling between the pump cavity mode and the Stokes cavity mode is explored in the presence of cavity radiation losses, linear material absorption, two-photon absorption, and free-carrier absorption. The refractive index shift from the Kerr effect, free-carrier dispersion, and thermal dispersion are also considered in the coupled-mode equations. These equations can be numerically integrated to describe the dynamical behavior of the system for the designed *L5* photonic band gap nanocavities. Specific examples such as lasing threshold, both pump and Stokes seed in continuous wave (CW) operation, pulsed pump with CW Stokes seed, on-off gain (loss), and the interaction of pump pulse and Stokes pulse are investigated in detail.

## 2. Design concept and coupled-mode theory

Stimulated Raman scattering is an inelastic two-photon process, where an incident photon interacts with an excited state of the material (the LO and TO phonons of single-crystal silicon). The strongest Stokes peak arises from single first-order Raman phonon (three-fold degenerate) at the Brillouin zone center. We have proposed a photonic band gap cavity with five linearly aligned missing air holes (*L5*) in an air-bridge triangular lattice photonic crystal slab with thickness of 0.6a and the radius of air holes is 0.29a, where the lattice period a = 420 nm [33]. The designed cavity supports two even modes, pump mode and Stokes mode, with spacing 15.6 THz, corresponding to the optical phonon frequency in monolithic silicon. Figure 1 shows the scanning electron micrograph (SEM) of the designed and fabricated *L5* cavity coupled with photonic crystal waveguide. Figure 2(a) and 2(b) show the electric field profile ($E_y$) at the middle of the slab for pump mode and Stokes mode calculated from 3D FDTD method.

Coupling between pump mode and Stokes mode in SRS can be understood classically with nonlinear polarizations $\mathbf{P}_{NL}^{(3)}$. The dynamics of SRS is governed through a set of time-dependent coupled nonlinear equations (in MKS) [49],

$$\nabla \times \nabla \times \mathbf{E}_p + \frac{\varepsilon_p}{\varepsilon_0 c^2} \frac{\partial^2 \mathbf{E}_p}{\partial t^2} = -\frac{1}{\varepsilon_0 c^2} \frac{\partial^2 \mathbf{P}_{NL}^{(3)}(\omega_p)}{\partial t^2} \qquad (1)$$

$$\nabla \times \nabla \times \mathbf{E}_S + \frac{\varepsilon_S}{\varepsilon_0 c^2} \frac{\partial^2 \mathbf{E}_S}{\partial t^2} = -\frac{1}{\varepsilon_0 c^2} \frac{\partial^2 \mathbf{P}_{NL}^{(3)}(\omega_S)}{\partial t^2} \qquad (2)$$

where $\mathbf{E}_p$ and $\mathbf{E}_S$ are the electric fields of pump mode and Stokes mode respectively, $\mathbf{P}_{NL}^{(3)}(\omega_p)$ and $\mathbf{P}_{NL}^{(3)}(\omega_S)$ are the third-order nonlinear polarizability, $\varepsilon_p$ and $\varepsilon_S$ are the dielectric constants. The third-order nonlinear polarization,

$$\mathbf{P}_{NL}^{(3)}(\omega_p) = 6\varepsilon_0 \chi_{ijkl}^{(3)}(\omega_p) \mathbf{E}_S \mathbf{E}_S^* \mathbf{E}_p \qquad (3)$$

$$\mathbf{P}_{NL}^{(3)}(\omega_S) = 6\varepsilon_0 \chi_{ijkl}^{(3)}(\omega_S) \mathbf{E}_p \mathbf{E}_p^* \mathbf{E}_S \qquad (4)$$

where $\chi_{ijkl}^{(3)}$ is the third-order nonlinear electric susceptibility.

Assume the electric fields of pump mode and Stokes mode are,

$$\mathbf{E}_p(\mathbf{r},t) = a_p(t) \frac{\mathbf{A}_p(\mathbf{r})}{\sqrt{N_p}} e^{-i\omega_p t} \qquad (5)$$

$$\mathbf{E}_S(\mathbf{r},t) = a_S(t) \frac{\mathbf{A}_S(\mathbf{r})}{\sqrt{N_S}} e^{-i\omega_S t} \qquad (6)$$

where $\mathbf{A}_p(r)$ and $\mathbf{A}_S(r)$ are the spatial part of the modes. $a_p(t)$ and $a_S(t)$ are slowly varying envelopes of the pumps and Stokes modes respectively. The amplitude is normalized by the spatial part to represent the energy of the mode (in units of Joules)

$$U_i = |a_i|^2 = \frac{1}{2}\int \varepsilon_i(\mathbf{r})|\mathbf{E}_i(\mathbf{r},t)|^2 d^3r, \text{ and } N_i = \frac{1}{2}\int \varepsilon_i(\mathbf{r})|\mathbf{A}_i(\mathbf{r})|^2 d^3r, \text{ i = p, S} \tag{7}$$

Then the third-order nonlinear polarization,

$$\mathbf{P}_{NL}^{(3)}(\omega_p) = 6\varepsilon_0 \chi_R^{(3)}(\omega_p)|a_S|^2 a_p \frac{|\mathbf{A}_S|^2 \mathbf{A}_p}{N_S \sqrt{N_p}} e^{-i\omega_p t} \tag{8}$$

$$\mathbf{P}_{NL}^{(3)}(\omega_S) = 6\varepsilon_0 \chi_R^{(3)}(\omega_S)|a_p|^2 a_S \frac{|\mathbf{A}_p|^2 \mathbf{A}_S}{N_p \sqrt{N_S}} e^{-i\omega_S t} \tag{9}$$

where $\chi_R^{(3)}$ is the effective third-order nonlinear electric susceptibility related to the polarization directions of $\mathbf{E}_{p,S}$ and $\mathbf{P}_{NL}^{(3)}$ [45],

$$\chi_R^{(3)} = \chi_{ijkl}^{(3)} \boldsymbol{\alpha}^* \boldsymbol{\beta} \boldsymbol{\gamma} \boldsymbol{\delta} \tag{10}$$

$\boldsymbol{\alpha}$ is the unit vector of the induced polarization $\mathbf{P}_{NL}^{(3)}$, $\boldsymbol{\beta}, \boldsymbol{\gamma}$ and $\boldsymbol{\delta}$ are unit vectors of the interacting fields $\mathbf{E}_{p,S}$. Next, by substituting the expression of $\mathbf{E}_{p,S}$ and $\mathbf{P}_{NL}^{(3)}$ into the coupled wave equations, taking the slowly varying envelope approximation $\left|\frac{\partial^2 a}{\partial t^2}\right| \ll \omega \left|\frac{\partial a}{\partial t}\right|$ and only the real part of amplitudes, we obtain

$$\frac{da_p}{dt}\mathbf{A}_p = -\frac{3\omega_p \text{ Im}(\chi_R^{(3)}(\omega_p))}{\varepsilon_p/\varepsilon_0}\left(\frac{|\mathbf{A}_S|^2}{N_S}\mathbf{A}_p\right)|a_S|^2 a_p \tag{11}$$

$$\frac{da_S}{dt}\mathbf{A}_S = -\frac{3\omega_S \text{ Im}(\chi_R^{(3)}(\omega_S))}{\varepsilon_S/\varepsilon_0}\left(\frac{|\mathbf{A}_p|^2}{N_p}\mathbf{A}_S\right)|a_p|^2 a_S \tag{12}$$

Then multiply the equations by the operator $\frac{1}{2}\int_{Si}\varepsilon_{p,S}(\mathbf{r})\mathbf{A}^*_{p,S}(\mathbf{r})d^3r$, for which $\int_{Si}$ only integrates over the silicon region of the photonic crystal cavity. This results in the coupled-mode rate equations, relating the pump and Stokes evolutions without any loss terms currently,

$$\frac{da_p}{dt} = -\left(\frac{\omega_p}{\omega_S}\right)g^c_S|a_S|^2 a_p \tag{13}$$

$$\frac{da_S}{dt} = g^c_S|a_p|^2 a_S \tag{14}$$

where the Raman gain coefficient in the photonic band gap nanocavities, $g^c_S$ [J$^{-1}$s$^{-1}$], is,

$$g^c_S = -\frac{6\omega_S \operatorname{Im}[\chi^{(3)}_R(\omega_S)]}{\varepsilon_0 n_p^2 n_S^2 V_R} \tag{15}$$

where $n_{p,S}$ are the refractive indices at the pump and Stokes wavelengths $\lambda_p$ and $\lambda_S$ respectively, and $n^2_{p,S} = \varepsilon_{p,S}/\varepsilon_0$. The Raman gain is assumed constant, without saturation or parametric instability. The effective modal volume $V_R$ for Raman scattering indicates the spatial overlap between the pump mode and the Stokes mode,

$$V_R = \frac{\int n_p^2(\mathbf{r})|\mathbf{A}_p(\mathbf{r})|^2 d^3r \cdot \int n_S^2(\mathbf{r})|\mathbf{A}_S(\mathbf{r})|^2 d^3r}{\int_{Si} n_p^2(\mathbf{r})|\mathbf{A}_p(\mathbf{r})|^2 \cdot n_S^2(\mathbf{r})|\mathbf{A}_S(\mathbf{r})|^2 d^3r} \tag{16}$$

The bulk gain coefficient, $g^B_R$ [m/W], is

$$g^B_R = -\frac{12\omega_S \operatorname{Im}[\chi^{(3)}_R(\omega_S)]}{\varepsilon_0 n_p n_S c^2} \tag{17}$$

so that

$$g_S^c = \left(\frac{c^2}{2n_p n_S V_R}\right) g_R^B \qquad (18)$$

Note that in this classical formulation, the Raman gain coefficient in the photonic band gap nanocavities $g_S^c$ is still equivalent to the bulk Raman gain coefficient $g_R^B$, since possible cavity quantum electrodynamics enhancements are not yet considered.

The electric field of input pump wave and input Stokes wave in the waveguide are,

$$\mathbf{E}_{i,in}(\mathbf{r},t) = s_i(t) \frac{\mathbf{S}_i(\mathbf{r})}{\sqrt{N_{i,in}}} e^{-i\omega_i t} \qquad (19)$$

where the field amplitude is normalized by $N_{i,in} = \frac{1}{2}\int n_{i,in}(\mathbf{r})\varepsilon_0 c |\mathbf{S}_i(\mathbf{r})|^2 d^2 r$ ($i = p, S$), to represent the input power $P_{i,in} = |s_i|^2 = \frac{1}{2}\int n_{i,in}(\mathbf{r})\varepsilon_0 c |\mathbf{E}_{i,in}(\mathbf{r},t)|^2 d^2 r$. Now, considering the in-plane waveguide coupling loss $1/\tau_{i,in}$ and the vertical radiation loss $1/\tau_{i,v}$ [36], the coupled-mode rate equations are

$$\frac{da_p}{dt} = -\frac{1}{2\tau_p} a_p - \left(\frac{\omega_p}{\omega_S}\right) g_S^c |a_S|^2 a_p + \kappa_p s_p \qquad (20)$$

$$\frac{da_S}{dt} = -\frac{1}{2\tau_S} a_S + g_S^c |a_p|^2 a_S + \kappa_S s_S \qquad (21)$$

where $1/\tau_i = 1/\tau_{i,in} + 1/\tau_{i,v}$, and $1/\tau_{i,in/v} = \omega_i/Q_{i,in/v}$, $1/\tau_i = \omega_i/Q_i$, ($i = p, S$). $1/\tau_{i,in}$ and $1/\tau_{i,v}$ are the loss rates into waveguide (in-plane) and into freespace (vertical). $\kappa_i$ is the coupling coefficient of input pump wave $s_p(t)$ or Stokes wave $s_S(t)$ coupled to the pump mode $a_p(t)$ or the Stokes mode $a_S(t)$ of the cavity, and $\kappa_i = \sqrt{1/\tau_{i,in}}$. The threshold pump power for the stimulated Raman lasing is obtained from Equations (20) and (21),

$$P_{in,th} = \frac{\pi^2 n_p n_S}{\lambda_p \lambda_S} \frac{V_R}{g_R^B} \left( \frac{Q_{p,in}}{Q_p^2 Q_S} \right) \quad (22)$$

The lasing threshold scales with $V_R/Q_p Q_S$ as illustrated in Equation (22). This therefore suggests the motivation for small $V_R$ cavities with high-$Q$ factors.

Now, considering the total loss rate $1/\tau_{i,total}$ and the shifted resonant frequency $\Delta\omega_i$ of pump mode and Stokes mode, the coupled-mode rate equations are therefore [40, 42]

$$\frac{da_p}{dt} = \left( -\frac{1}{2\tau_{p,total}} + i\Delta\omega_p \right) a_p - \left( \frac{\omega_p}{\omega_S} \right) g_S^c |a_S|^2 a_p + \kappa_p s_p \quad (23)$$

$$\frac{da_S}{dt} = \left( -\frac{1}{2\tau_{S,total}} + i\Delta\omega_S \right) a_S + g_S^c |a_p|^2 a_S + \kappa_S s_S \quad (24)$$

This framework has been reported earlier in Johnson et al. [40] and Uesugi et al. [42] for a single frequency in cavities. We further advance these investigations for the pump-Stokes interactions, as well as studying the lasing thresholds and dynamics under various conditions. The total loss rate for each cavity mode is:

$$1/\tau_{i,total} = 1/\tau_{i,in} + 1/\tau_{i,v} + 1/\tau_{i,lin} + 1/\tau_{i,TPA} + 1/\tau_{i,FCA} \quad (25)$$

The linear material absorption $1/\tau_{lin}$ is assumed small since operation is within the bandgap of the silicon material. $1/\tau_{TPA}$ and $1/\tau_{FCA}$ are the loss rates due to two-photon absorption (TPA) and free-carrier absorption (FCA) respectively. The modal-averaged TPA loss rates are [28],

$$\frac{1}{\tau_{p,TPA}} = \frac{\beta_{Si} c^2}{n_p^2 V_{p,TPA}} |a_p|^2 + \frac{\beta_{Si} c^2}{n_p^2 V_{o,TPA}} 2|a_S|^2 \quad (26)$$

$$\frac{1}{\tau_{S,TPA}} = \frac{\beta_{Si}c^2}{n_S^2 V_{S,TPA}}|a_S|^2 + \frac{\beta_{Si}c^2}{n_S^2 V_{o,TPA}}2|a_p|^2 \qquad (27)$$

where the first terms represent the TPA due to two pump photons or two Stokes photons. The second terms represent that one pump photon and one Stokes photon are absorbed simultaneously. $\beta_{Si}$ is the TPA coefficient of bulk silicon. The effective mode volume for TPA, $V_{i,TPA}$, is

$$V_{i,TPA} = \frac{\left(\int n_i^2(\mathbf{r})|\mathbf{A}_i(\mathbf{r})|^2 dr^3\right)^2}{\int_{Si} n_i^4(\mathbf{r})|\mathbf{A}_i(\mathbf{r})|^4 dr^3} \qquad (28)$$

$V_{o,TPA}$ indicates the spatial overlap between the pump mode and the Stokes mode, and $V_{o,TPA} = V_R$. We note that the ($\beta_{Si}/V_{i,TPA}$) TPA term is the effective silicon-air contribution, summing over the cavity modal distributions within the solid and negligible contribution from air, and neglecting crystal anisotropy. The bulk TPA coefficient is also assumed to be frequency-independent [50], and without surface modification for simplicity. The modal-averaged FCA loss rates are,

$$\frac{1}{\tau_{i,FCA}} = \frac{c}{n_i}\alpha_{i,FCA} = \frac{c}{n_i}(\sigma_{i,e} + \sigma_{i,h})N(t) \qquad (29)$$

From the Drude model [51], the absorption cross-sections for electrons and holes, $\sigma_{i,e/h}$, is

$$\sigma_{i,e/h} = \frac{e^2}{cn_i\omega_i^2\varepsilon_0 m_{e/h}^*\tau_{relax,e/h}} \qquad (30)$$

Here $e$ is the electron charge, $\tau_{relax,e/h}$ the relaxation time of carriers, and $m^*_{e/h}$ the effective mass of carriers. The mode-averaged free carriers density (electron-hole pairs) generated by TPA is $N(t)$, which is governed by the rate equation [40],

$$\frac{dN}{dt} = -\frac{N}{\tau_{fc}} + G \tag{31}$$

The mode-averaged generation rate of free-carriers $G$ can be calculated from the mode-averaged TPA loss rate.

$$G = \frac{\beta_{Si} c^2}{2\hbar\omega_p n_p^2 V_{p,FCA}^2}|a_p|^4 + \frac{\beta_{Si} c^2}{2\hbar\omega_S n_S^2 V_{S,FCA}^2}|a_S|^4$$

$$+ \frac{\beta_{Si} c^2}{\hbar(\omega_p + \omega_S) n_p^2 V_{Sp,FCA}^2} 2|a_S|^2|a_p|^2 + \frac{\beta_{Si} c^2}{\hbar(\omega_p + \omega_S) n_S^2 V_{pS,FCA}^2} 2|a_p|^2|a_S|^2 \tag{32}$$

The expressions of effective mode volume for FCA, $V_{FCA}$, are

$$V_{i,FCA}^2 = \frac{\left(\int n_i^2(\mathbf{r})|\mathbf{A}_i(\mathbf{r})|^2 dr^3\right)^3}{\int_{Si} n_i^6(\mathbf{r})|\mathbf{A}_i(\mathbf{r})|^6 dr^3} \tag{33}$$

$$V_{Sp,FCA}^2 = \frac{\left(\int n_p^2(\mathbf{r})|\mathbf{A}_p(\mathbf{r})|^2 dr^3\right)^2 \left(\int n_S^2(\mathbf{r})|\mathbf{A}_S(\mathbf{r})|^2 dr^3\right)}{\int_{Si} n_p^4(\mathbf{r})|\mathbf{A}_p(\mathbf{r})|^4 \cdot n_S^2(\mathbf{r})|\mathbf{A}_S(\mathbf{r})|^2 dr^3} \tag{34}$$

$$V_{pS,FCA}^2 = \frac{\left(\int n_S^2(\mathbf{r})|\mathbf{A}_S(\mathbf{r})|^2 dr^3\right)^2 \left(\int n_p^2(\mathbf{r})|\mathbf{A}_p(\mathbf{r})|^2 dr^3\right)}{\int_{Si} n_S^4(\mathbf{r})|\mathbf{A}_S(\mathbf{r})|^4 \cdot n_p^2(\mathbf{r})|\mathbf{A}_p(\mathbf{r})|^2 dr^3} \tag{35}$$

$\tau_{fc}$ is the effective free-carrier lifetime accounting for both recombination and diffusion. Time constants of radiative and Auger recombination, as well as from bulk defects and impurities, are assumed to be significantly slower than the free-carrier recombination and diffusion lifetime [50]. We note that while free-carrier lifetime can vary with carrier density and carrier density can vary spatially with intensity in the cavity, an effective lifetime is used here for simplicity. A quiescent carrier density of $N_0 = 10^{22}$ m$^{-3}$ is used in the initial condition for silicon.

In equations (23) and (24), $\Delta\omega_i$ is the detuning of the resonance frequency of the cavity from the input light frequency due to the Kerr effect, free-carrier dispersion (FCD), and thermal dispersion. $\Delta\omega_i = \omega_i' - \omega_i$, $\omega_i'$ is the shifted resonant frequency of the cavity and $\omega_i$ is the input light frequency in the waveguide. Under first-order perturbation, the detuning of the resonance frequency can be expressed as [28]

$$\frac{\Delta\omega_i}{\omega_i} = -\frac{\Delta n_i}{n_i} = -\left(\frac{\Delta n_{i,Kerr}}{n_i} + \frac{\Delta n_{i,FCD}}{n_i} + \frac{\Delta n_{i,th}}{n_i}\right) \tag{36}$$

The detuning due to Kerr effect is

$$\frac{\Delta n_{p,Kerr}}{n_p} = \frac{cn_2}{n_p^2 V_{p,Kerr}}|a_p|^2 + \frac{cn_2}{n_p^2 V_{o,Kerr}}2|a_S|^2 \tag{37}$$

$$\frac{\Delta n_{S,Kerr}}{n_S} = \frac{cn_2}{n_S^2 V_{S,Kerr}}|a_S|^2 + \frac{cn_2}{n_S^2 V_{o,Kerr}}2|a_p|^2 \tag{38}$$

where the effective modal volume for Kerr effects $V_{i,Kerr} = V_{i,TPA}$ and $V_{o,Kerr} = V_{o,TPA}$. The first terms represent self-phase modulation and the second terms represent cross-phase modulation. The detuning due to free-carrier dispersion is related by

$$\frac{\Delta n_{i,FCD}}{n_i} = -\frac{1}{n_i}(\zeta_{i,e} + \zeta_{i,h})N(t) \tag{39}$$

From the Drude model [51], the material parameter with units of volume, $\zeta_{i,e/h}$, is

$$\zeta_{i,e/h} = \frac{e^2}{2n_i \omega_i^2 \varepsilon_0 m^*_{e/h}} \tag{40}$$

$$\frac{\Delta n_{i,th}}{n_i} = \frac{1}{n_i}\frac{dn_i}{dT}\Delta T \tag{41}$$

The mode-averaged temperature difference between the photonic crystal cavity and its environment $\Delta T$ is governed by [40]

$$\frac{d\Delta T}{dt} = -\frac{\Delta T}{\tau_{th}} + \frac{P_{abs}}{\rho_{Si} c_{p,Si} V_{cavity}} \tag{42}$$

where $\rho_{Si}$, $c_{p,Si}$ and $V_{cavity}$ are the density of silicon, the constant-pressure specific heat capacity of silicon and the volume of cavity respectively. The temperature decay life-time $\tau_{th}$ is determined by the thermal resistance $R$ of the air-bridge silicon photonic crystal cavities.

$$\tau_{th} = \rho_{Si} c_{p,Si} V_{cavity} R \tag{43}$$

The total absorbed power is given by

$$P_{abs} = P_{p,abs} + P_{S,abs} + P_{R,abs} \tag{44}$$

$$P_{i,abs} = \left(1/\tau_{i,lin} + 1/\tau_{i,TPA} + 1/\tau_{i,FCA}\right)|a_i|^2 \tag{45}$$

Absorbed power due to Raman scattering generated optical phonon, $P_{R,abs}$, is

$$P_{R,abs} = 2(\omega_p/\omega_S - 1)g_S^c |a_p|^2 |a_S|^2 \tag{46}$$

Equations (23-25), (31), (36), and (42) therefore describe the dynamic behavior of SRS in photonic crystal nanocavities, and is numerically integrated in our work to describe the dynamical behavior of pump-Stokes interactions in our *L5* photonic crystal cavity system that supports the desired two-mode frequencies at the appropriate LO/TO phonon spacing.

## 3. Numerical analysis

### 3.1 Lasing threshold

We now consider the case of lasing threshold. Around the lasing threshold, the Stokes gain equals the losses, and the Stokes mode energy $|a_S|^2$ is much smaller than the pump mode energy $|a_p|^2$, equation (24) is simplified to

$$g_S^c |a_p|_{th}^2 = \frac{1}{2\tau_{S,total}} \tag{47}$$

The loss rate due to TPA of pump mode is

$$\frac{1}{\tau_{S,TPA}} = \frac{\beta_{Si} c^2}{n_S^2 V_{o,TPA}} 2|a_p|_{th}^2 \tag{48}$$

The free carriers generated by TPA of pump mode is

$$N = \tau_{fc} G \tag{49}$$

$$G = \frac{\beta_{Si} c^2}{2\hbar \omega_p n_p^2 V_{p,FCA}^2} |a_p|_{th}^4 \tag{50}$$

Figure 3 shows the threshold pump mode energy $|a_p|^2_{th}$ as a function of $Q_S$ for different free-carrier lifetimes $\tau_{fc}$. All the parameters used in calculation are presented in Table 1. By comparing the curve in the absence of TPA and FCA ($\beta_{Si} = 0$) and the curve with TPA but without FCA ($\tau_{fc} = 0$), it is observed that TPA increases the threshold pump mode energy but the effect of TPA is relatively weak. The effect of TPA-induced FCA is much more dramatic as shown for different free-carrier lifetimes $\tau_{fc}$. The lasing threshold increases when $\tau_{fc}$ is larger. There is a minimum Stokes $Q_S$ required for lasing, as seen in the solutions plotted in Figure 3. If $Q_S$ is lower than a critical value for certain $\tau_{fc}$, there is no solution numerically and physically this translates to an absence of a lasing threshold regardless of the pump intensity. For increasing $\tau_{fc}$, the critical value of $Q_S$ increases monotonically as can be seen in Figure 3. The solid and dotted curves show the lasing and shutdown thresholds, respectively [44]. The shutdown threshold is the pump power in which the lasing output power returns to zero due to increasing TPA and FCA. For the *L5* cavity studied in the present paper, $Q_S$ = 21,000, the maximum $\tau_{fc}$ is around 0.175 ns, and the threshold pump mode energy is 29 fJ. For air-bridged silicon photonic band gap nanocavities, $\tau_{fc}$ = 0.5 ns [42], which is much higher than the maximum $\tau_{fc}$. In order to get lasing for this cavity, instead of using CW pump signal, pulse pump signal can be used to reduce the TPA-induced FCA for loss reduction.

We now solve for the input-output characteristics of photonic band gap defect cavity laser by considering equation (24) in steady state. Figure 4 shows the laser input-output characteristics with different free-carrier lifetimes $\tau_{fc}$. $Q_S$ = 30,000 and $Q_S$ = 60,000 are considered for comparison. Cavities with higher $Q_S$ have higher output Stokes signals, lower lasing threshold pump mode energies and higher shutdown thresholds. The required corresponding pump power in the input waveguide can also be calculated based on equation (23) in steady state.

**3.2 Lasing dynamics**

We now consider the dynamics of the Raman lasing interactions. Equations (23-25), (31), (36), and (42) are numerically integrated with a variable order Adams-Bashforth-Moulton predictor-corrector method (*Matlab*® ode113 solver). All the parameters used in calculation are presented in Table 1.

Figure 5 shows the dynamics of Raman amplification with 60 mW CW pump wave and 10 μW CW Stokes seed signal, free carrier lifetime is 0.5 ns. In the beginning when free carrier density is low, Raman gain is greater than loss and there is amplification. When free carrier density increases, FCA dominates the loss and Stokes signal is suppressed. The temperature difference then increases significantly. The cavity resonance is red shifted and the pump mode energy goes down. From the numerical results, the Kerr effect is predominantly weak. The FCD effect dominates at first when the temperature difference is low, with a resulting blue shift. Eventually thermal effect dominates, with a resulting red shift. Consider the case of a different carrier lifetime at 0.1 ns. Figure 6 shows the calculated results with free carrier lifetime of 0.1 ns. With lower free carrier lifetime, the free carrier density and the temperature difference are lower, so that the net Raman gain is greater than zero. The oscillation of Stokes mode energy near $t = 0.5$ ns is due to the dispersion induced Stokes resonance frequency shift.

In order to get lasing from this cavity, instead of using CW pump signal, pulse pump signal with pulse width narrower than the free carrier lifetime is used to reduce the TPA-induced FCA, so as to reduce loss and increase net gain. Figure 7 shows the dynamics of Raman amplification with pump pulse of 60 mW peak power, pulse width $T_{FWHM} = 50$ ps and 10 µW CW Stokes signal. The free carrier density and the temperature difference are significantly reduced, and a strong Stokes pulse is generated by the pump pulse. The resonance frequency shift is also significantly reduced by the pump pulse operation.

Consider now the case of on-off gain and loss in our cavity system, where the probe signal changes between the pump pulse on and off when the probe frequency is on- (off-) resonance with the Stokes frequency [9]. Pulsed pump beam with 60 mW peak power and $T_{FWHM} = 50$ ps, and CW probe beam with 1 mW power are used. Figure 8 shows the on-off gain and on-off loss. When the probe is on the Stokes frequency, an increase in the probe signal due to the SRS is observed, and the on-off gain is around 8 dB. When the probe frequency is detuned from the Stokes frequency (no Raman gain), the loss in the probe signal due to the pump pulse generated free carriers is observed, and the on-off loss is around 20 dB. Figures 9 and 10 show the dynamics of the system.

We also consider the interaction of both pump and Stokes pulses with comparable peak power, with numerical results shown in Figure 11. Pump peak power is 60 mW and Stokes peak power is 20 mW. Both pump pulse and Stokes pulse have $T_{FWHM}$ = 50 ps. The Stokes pulse is amplified by the pump pulse. Due to the high free carrier density induced FCD effect and high $Q_S$, there is an observed oscillation in the amplified Stokes pulse.

## 4. Conclusions

In this work we have derived the coupled-mode equations for stimulated Raman scattering in high-$Q/V_m$ silicon photonic band gap nanocavities towards optically-pumped silicon lasing. Both the lasing threshold and the lasing dynamics are numerically studied in the presence of cavity radiation losses, linear material absorption, two-photon absorption, and free-carrier absorption, together with the refractive index shift from the Kerr effect, free-carrier dispersion, and thermal dispersion. With increasing the cavity $Q$ factors and decreasing the free carrier lifetimes, the reduction in the threshold pump energy is solved numerically, considering all mechanisms and realistic conditions. With CW pump operation, the Stokes signal is suppressed due to strong TPA-induced FCA. With pulse pump operation, the TPA-induced FCA is significantly reduced and Stokes net gain increases, which shows that compact Raman amplifiers and lasers based on high-$Q/V_m$ silicon photonic band gap nanocavities are feasible.

## Acknowledgement

The authors thank Xiaogang Chen for helpful discussions. This work was partially supported by Columbia University Initiatives in Science and Engineering for Nanophotonics.

**Table 1. Parameters used in coupled-mode theory**

**Fig. 1.** SEM of *L5* cavity coupled with photonic crystal waveguide.

**Fig. 2.** The electric field profile (*E$_y$*) of pump mode (a) and Stokes mode (b).

**Fig. 3.** Threshold pump mode energy versus *Q$_S$* of *L5* cavity for different values of free-carrier lifetimes $\tau_{fc}$, the solid curve and dotted curve show the lasing and shutdown thresholds, respectively.

**Fig. 4.** Input-output characteristics of photonic band gap cavity laser for different values of free-carrier lifetimes $\tau_{fc}$, the solid curve and dashed curve correspond to *Q$_S$* = 30,000 and *Q$_S$* = 60,000, respectively.

**Fig. 5.** Dynamics of Raman amplification with 60 mW CW pump wave and 10 μW CW Stokes seed signal, $\tau_{fc}$ is 0.5 ns.

**Fig. 6.** Dynamics of Raman amplification with 60 mW CW pump wave and 10 μW CW Stokes seed signal, $\tau_{fc}$ is 0.1 ns.

**Fig. 7.** Dynamics of Raman amplification with pulse pump of 60 mW peak power, $T_{FWHM}$ = 50 ps and 10 μW CW Stokes signal, $\tau_{fc}$ is 0.5 ns.

**Fig. 8.** Raman on-off gain and on-off loss with pump pulse of 60 mW peak power, $T_{FWHM}$ = 50 ps and 1 mW CW probe signal, $\tau_{fc}$ is 0.5 ns.

**Fig. 9.** Dynamics of Raman on-off gain with pump pulse of 60 mW peak power, $T_{FWHM}$ = 50 ps and 1 mW CW probe signal, $\tau_{fc}$ is 0.5 ns.

**Fig. 10.** Dynamics of Raman on-off loss with pulse pump of 60 mW peak power, $T_{FWHM}$ = 50 ps and 1 mW CW probe signal, $\tau_{fc}$ is 0.5 ns.

**Fig. 11.** Dynamics of Raman interaction of pump pulse with 60 mW peak power and Stokes pulse with 20 mW peak power, $T_{FWHM}$ = 50 ps, $\tau_{fc}$ is 0.5 ns.

**Table 1. Parameters used in coupled-mode theory**

| Parameter | Symbol | Value | Source |
|---|---|---|---|
| Refractive index of silicon | $n_i$ | 3.485 | [52] |
| Wavelength of pump mode | $\lambda_p$ | 1496.7 nm | FDTD |
| Wavelength of Stokes mode | $\lambda_S$ | 1623.1 nm | FDTD |
| Pump mode in-plane $Q$ | $Q_{p,in}$ | 960 | FDTD |
| Pump mode vertical $Q$ | $Q_{p,v}$ | 960 | FDTD |
| Stokes mode in-plane $Q$ | $Q_{S,in}$ | 42,000 | FDTD |
| Stokes mode vertical $Q$ | $Q_{S,v}$ | 42,000 | FDTD |
| Linear material absorption loss | $1/\tau_{lin}$ | 0.86 GHz | [40] |
| Raman mode volume | $V_R$ | 0.544868 x 10$^{-18}$ m$^3$ | FDTD |
| TPA mode volume of pump mode | $V_{p,TPA}$ | 0.258387 x 10$^{-18}$ m$^3$ | FDTD |
| TPA mode volume of Stokes mode | $V_{S,TPA}$ | 0.396806 x 10$^{-18}$ m$^3$ | FDTD |
| FCA mode volume of pump mode | $V_{p,FCA}$ | 0.202601 x 10$^{-18}$ m$^3$ | FDTD |
| FCA mode volume of Stokes mode | $V_{S,FCA}$ | 0.299289 x 10$^{-18}$ m$^3$ | FDTD |
| FCA mode volume | $V_{Sp,FCA}$ | 0.337575 x 10$^{-18}$ m$^3$ | FDTD |
| FCA mode volume | $V_{pS,FCA}$ | 0.368572 x 10$^{-18}$ m$^3$ | FDTD |
| Bulk Raman gain coefficient | $g_R^B$ | 2.9 x 10$^{-10}$ m/W | [7] |
| TPA coefficient | $\beta_{Si}$ | 4.4 x 10$^{-12}$ m/W | [6] |
| Kerr coefficient | $n_2$ | 4.4 x 10$^{-18}$ m$^2$/W | [53] |
| Free-carrier lifetime | $\tau_{fc}$ | 0.5 ns | [28, 42] |
| Absorption cross-sections for electrons | $\sigma_{i,e}$ | 8.5 x 10$^{-22}$ m$^2$ | [51, 54] |
| Absorption cross-sections for holes | $\sigma_{i,h}$ | 6.0 x 10$^{-22}$ m$^2$ | [51, 54] |
| FCD parameter for electrons | $\zeta_{i,e}$ | 8.8 x 10$^{-28}$ m$^3$ | [51, 54] |
| FCD parameter for holes | $\zeta_{i,h}$ | 4.6 x 10$^{-28}$ m$^3$ | [51, 54] |
| Density of silicon | $\rho_{Si}$ | 2.33 x 10$^3$ kg/m$^3$ | [55] |
| Constant-pressure specific heat capacity of silicon | $c_{p,Si}$ | 0.7 x 10$^3$ J·kg$^{-1}$·K$^{-1}$ | [55] |
| Volume of cavity | $V_{cavity}$ | 0.462 x 10$^{-18}$ m$^3$ | ~ L x W x H |
| Thermal resistance | $R$ | 50 K/mW | [42, 56] |
| Temperature dependence of refractive index | $dn_i/dT$ | 1.85 x 10$^{-4}$ K$^{-1}$ | [57] |

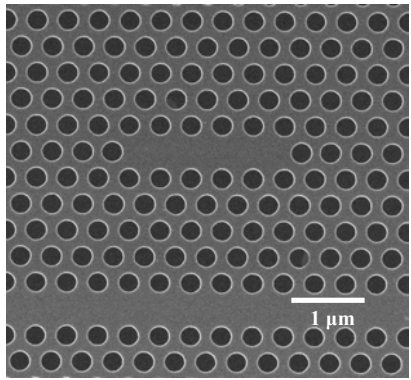

**Fig. 1. SEM picture of *L5* cavity coupled with photonic crystal waveguide.**

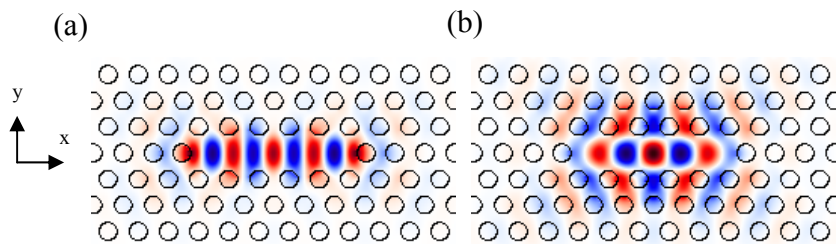

**Fig. 2. The electric field profile ($E_y$) of pump mode (a) and Stokes mode (b).**

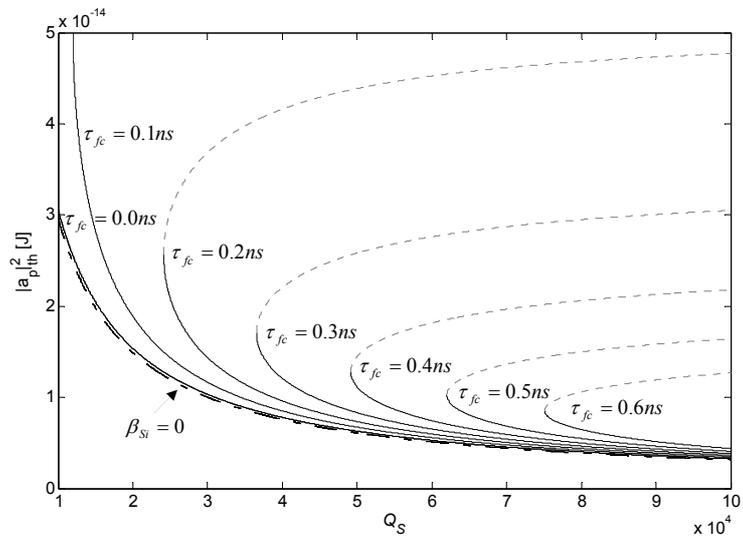

**Fig. 3. Threshold pump mode energy versus $Q_S$ of *L5* cavity for different values of free-carrier lifetimes $\tau_{fc}$, the solid curve and dotted curve show the lasing and shutdown thresholds, respectively.**

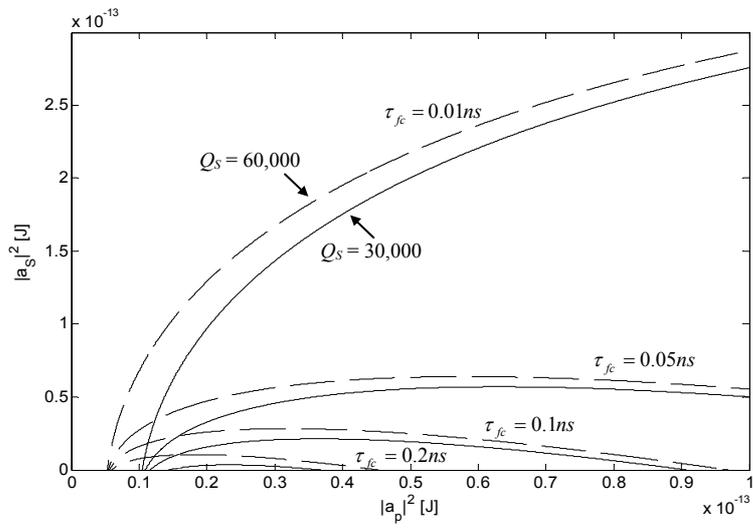

Fig. 4. Input-output characteristics of photonic band gap defect cavity laser for different values of free-carrier lifetimes $\tau_{fc}$, the solid curve and dashed curve correspond to $Q_S = 30,000$ and $Q_S = 60,000$, respectively.

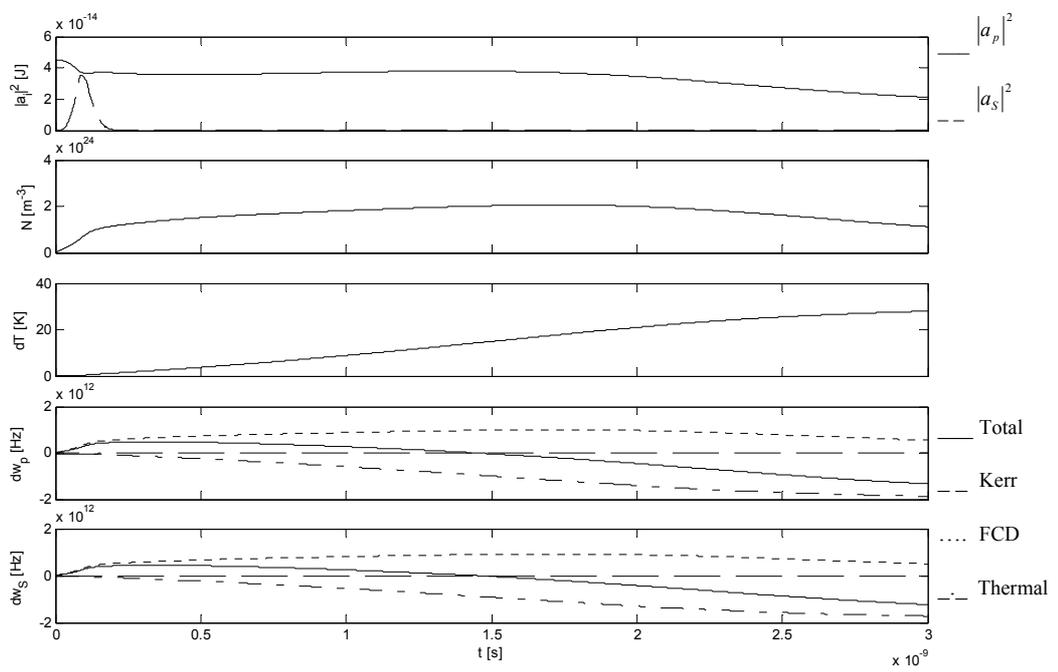

**Fig. 5. Dynamics of Raman amplification with 60 mW CW pump wave and 10 µW CW Stokes seed signal, $\tau_{fc}$ is 0.5 ns.**

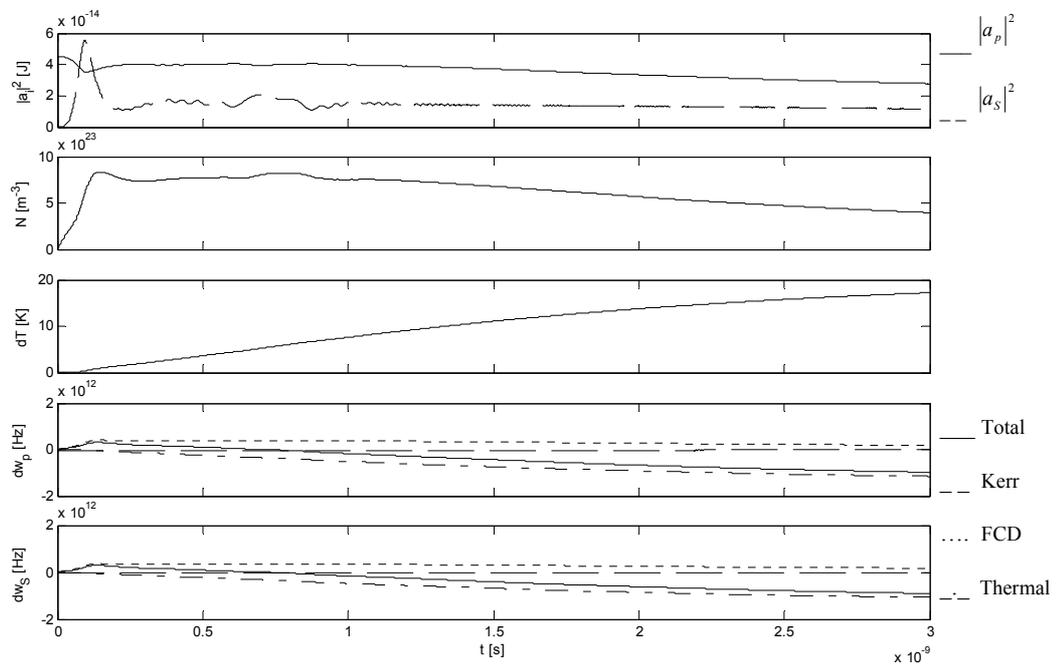

**Fig. 6. Dynamics of Raman amplification with 60 mW CW pump wave and 10 µW CW Stokes seed signal, $\tau_{fc}$ is 0.1 ns.**

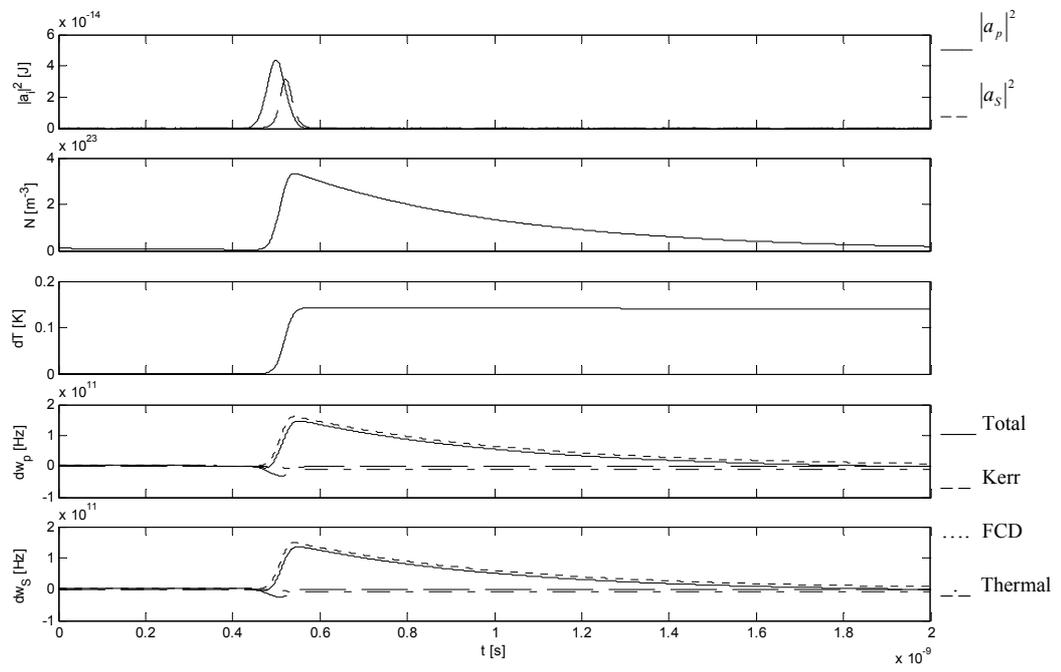

**Fig. 7. Dynamics of Raman amplification with pulse pump of 60 mW peak power, $T_{FWHM}$ = 50 ps and 10 µW CW Stokes signal, $\tau_{fc}$ is 0.5 ns.**

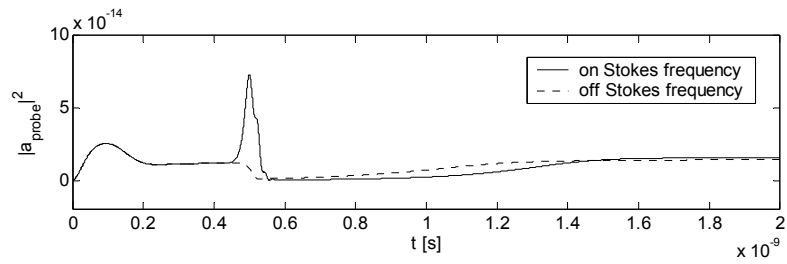

**Fig. 8. Raman on-off gain and on-off loss with pump pulse of 60 mW peak power, $T_{FWHM}$ = 50 ps and 1 mW CW probe signal, $\tau_{fc}$ is 0.5 ns.**

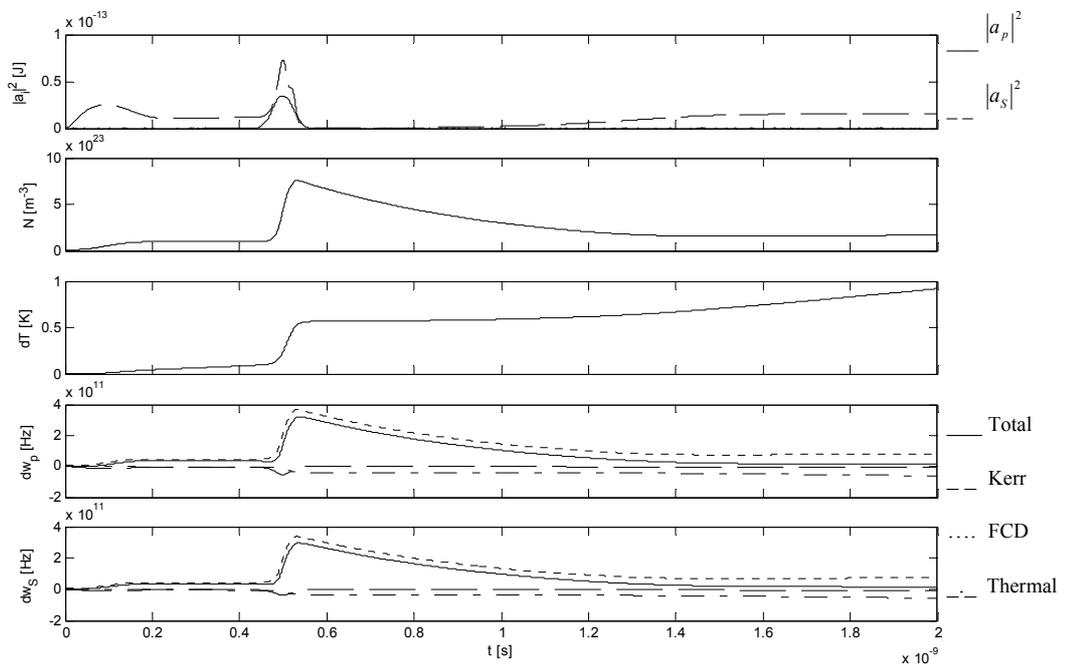

**Fig. 9. Dynamics of Raman on-off gain with pump pulse of 60 mW peak power, $T_{FWHM}$ = 50 ps and 1 mW CW probe signal, $\tau_{fc}$ is 0.5 ns.**

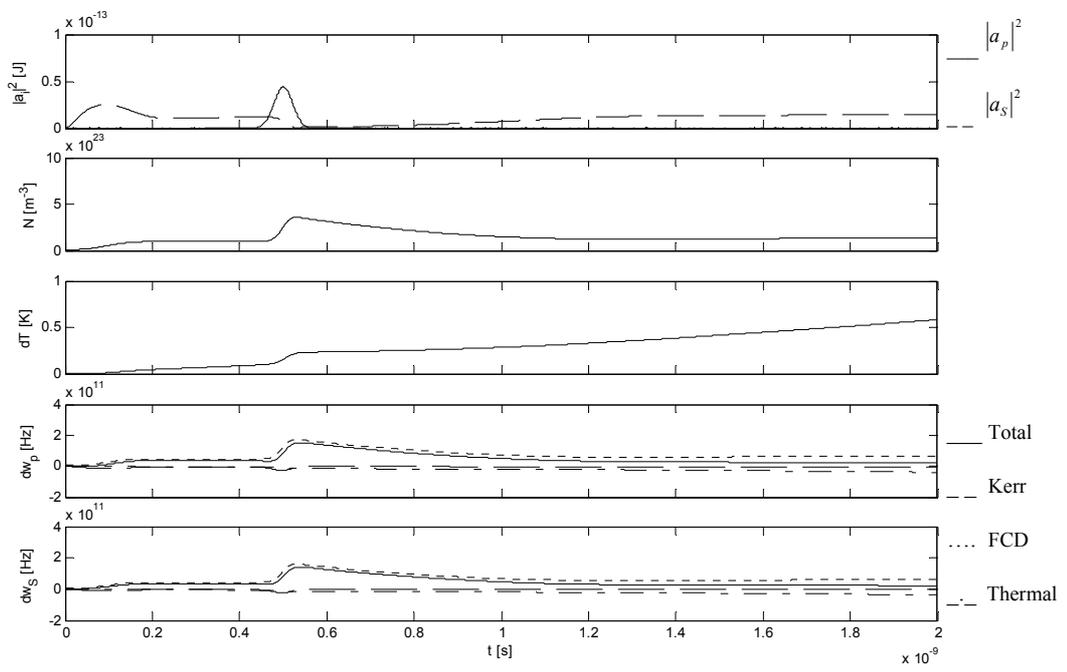

**Fig. 10. Dynamics of Raman on-off loss with pulse pump of 60 mW peak power,**

$T_{FWHM}$ **= 50 ps and 1 mW CW probe signal,** $\tau_{fc}$ **is 0.5 ns.**

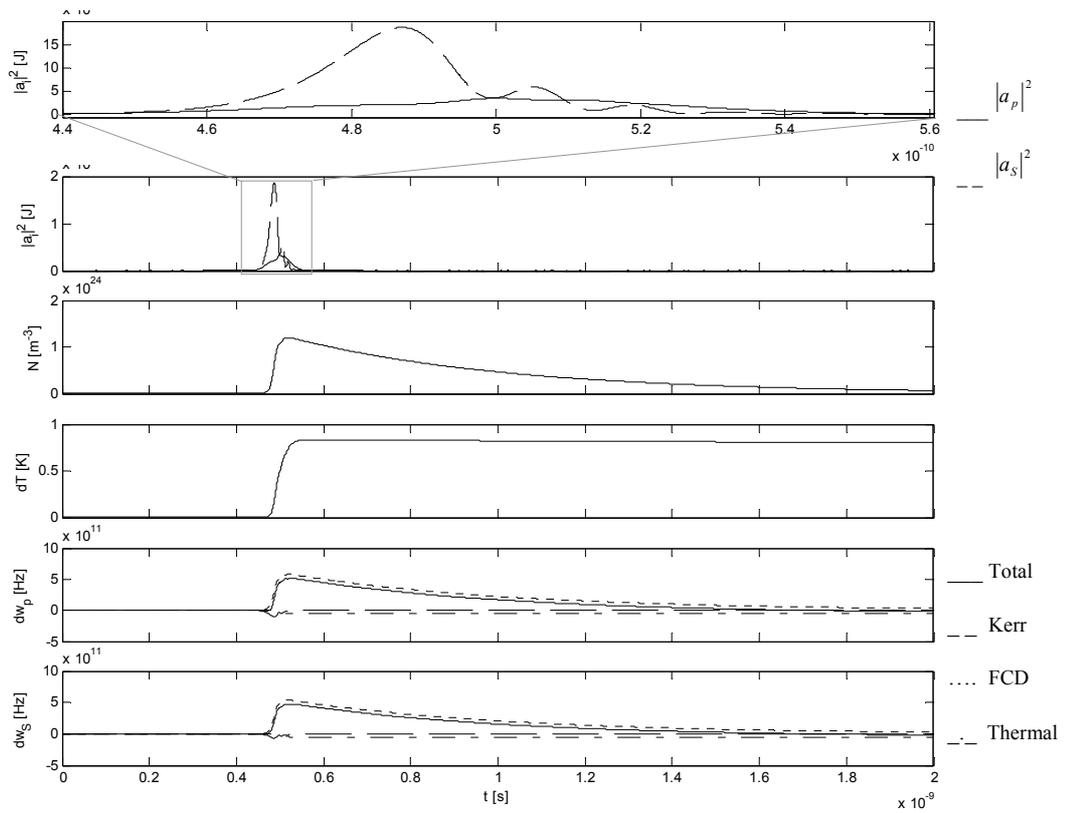

**Fig. 11. Dynamics of Raman interaction of pump pulse with 60 mW peak power and Stokes pulse with 20 mW peak power, $T_{FWHM}$ = 50 ps, $\tau_{fc}$ is 0.5 ns.**